\documentclass[prd,a4paper,superscriptaddress,aps,showpacs,twocolumn,floats,nofootinbib]{revtex4}
\usepackage{epsfig}
\usepackage{graphicx}
\usepackage{amsmath}
\usepackage{amssymb}
\usepackage{color}
\usepackage{soul}

\newcommand{\bx}{{\mathbf x}}

\newcommand{\RR}{{\cal R}}

\newcommand{\al}{\alpha}

\newcommand{\ep}{\epsilon}

\newcommand{\La}{\Lambda}
\newcommand{\la}{\lambda}
\newcommand{\Om}{\Omega}

\newcommand{\si}{\sigma}

\newcommand{\be}{\begin{equation}}
\newcommand{\ee}{\end{equation}}

\newcommand{\bea}{\begin{eqnarray}}
\newcommand{\eea}{\end{eqnarray}}
\newcommand{\bean}{\begin{eqnarray*}}
\newcommand{\eean}{\end{eqnarray*}}
\newcommand{\dd}{\partial}

\newcommand{\HH}{{\ensuremath{\mathcal H}}}

\newcommand{\FF}{\mathcal{F}}

\begin{document}

\title{Distance--redshift relation in plane symmetric universes} 

\author{Julian Adamek}
\email{julian.adamek@unige.ch}

\affiliation{D\'epartement de Physique Th\'eorique \& Center for Astroparticle Physics, Universit\'e de Gen\`eve,\\ 
24 quai Ernest Ansermet, 1211 Gen\`eve 4, Switzerland}

\author{Enea Di Dio}
\email{enea.didio@unige.ch}

\affiliation{D\'epartement de Physique Th\'eorique \& Center for Astroparticle Physics, Universit\'e de Gen\`eve,\\ 
24 quai Ernest Ansermet, 1211 Gen\`eve 4, Switzerland}

\author{Ruth Durrer}
\email{ruth.durrer@unige.ch}

\affiliation{D\'epartement de Physique Th\'eorique \& Center for Astroparticle Physics, Universit\'e de Gen\`eve,\\ 
24 quai Ernest Ansermet, 1211 Gen\`eve 4, Switzerland}

\author{Martin Kunz}
\email{martin.kunz@unige.ch}

\affiliation{D\'epartement de Physique Th\'eorique \& Center for Astroparticle Physics, Universit\'e de Gen\`eve,\\ 
24 quai Ernest Ansermet, 1211 Gen\`eve 4, Switzerland}

\affiliation{African Institute for Mathematical Sciences, 6 Melrose Road, Muizenberg 7945, South Africa}

\date{\today}

\begin{abstract}
Distance measurements are usually thought to probe the `background' metric of the universe, but in reality the presence
of perturbations will lead to deviations from the result expected in an exactly homogeneous and isotropic universe. At least 
in principle the presence of perturbations could even explain the observed distance--redshift relation without the need for
dark energy. In this paper we re-investigate a toy model where perturbations are plane symmetric,
and for which exact solutions are known in the fluid limit. However, if perturbations are large, shell-crossing occurs and
the fluid approximation breaks down. This prevents the study of the most interesting cases.
Here we use a general-relativistic $N$-body simulation that does not suffer from this
problem and which allows us to go beyond previous works. We show that even for very large plane-symmetric perturbations
we are not able to mimic the observed distance-redshift relation. We also discuss how the synchronous comoving gauge breaks
down when shell-crossing occurs, while metric perturbations in the longitudinal gauge remain small. For this
reason the longitudinal (Newtonian) gauge appears superior for relativistic $N$-body simulations of large-scale structure formation.
\end{abstract}


\maketitle

\vspace{0.1cm}
\section{Introduction}
The Physics Nobel Prize  2011 has been given ``for the discovery of the accelerated expansion of the Universe''~\cite{SN1a,SNnew}.
Interpreting this finding within the model of a Friedmann--Lema\^\i tre (FL) universe requires that the energy density of the Universe
is presently dominated by a component with strongly negative pressure, $p=w\rho$ with $w\sim -1$. The nature of this so-called
Dark Energy (DE) remains largely unexplained to date, and is considered as one of the grand challenges of cosmology.
In the cosmological standard model it is addressed by introducing a cosmological constant $\La$ which is equivalent to a
vacuum energy $\rho_\La =\La/(8\pi G)$ and obeys the equation of state $p_\La = -\rho_\La$. However, the required value of this
vacuum energy is considered unnatural, as it is much smaller than all known fundamental scales in particle physics and there is
no established mechanism which protects it from large quantum corrections. Therefore, although formally possible, this explanation
of DE remains unsatisfactory.

So far, the measurements pointing to the existence of DE rely mainly on the
distance--redshift relation which is valid in an FL universe~\cite{whatDE}. Independent measurements of, e.g. the expansion rate $H(z)$
are underway, but at present they are still relatively weak, see, e.g.,~\cite{arXiv:1108.2637}. In the future however, for instance
with the Euclid satellite\footnote{{\tt http://www.euclid-ec.org}}~\cite{Euclid}, it will be possible to measure $H(z)$
and the luminosity distance $d_L(z)$ independently to sufficient accuracy to test whether they obey the relation predicted in an
FL universe~\cite{MD}. 

There are many ways to address the DE problem. Most of them can be classified either as `Dark Energy' or `Dark Gravity' by
specifying whether they modify the right hand side of Einstein's equation, by introducing a new contribution to the energy momentum
tensor, or its left hand side by modifying the laws of gravity, making gravity weaker on large scales. For a review see,
e.g.~\cite{RR}.

However, there is also the `coincidence problem': why did DE start to become important roughly at the time when non-linear
structures have formed? This leads to the question whether one might be completely misled by using the distance--redshift
relation of a homogeneous FL universe when actually the true Universe is lumpy, inhomogeneous. If this were true, DE does not
exist but is inferred from an oversimplified, and hence inappropriate interpretation of the data. This would certainly be the most
conservative solution to the DE problem, requiring no new physics at all (although it does not explain why vacuum energy is
not large, as there is no obvious reason why zero should be a preferred value).

One might argue that on large scales over- and underdensities compensate and the distance--redshift relation is similar to the one
in an FL universe. However, since General Relativity is non-linear, the relation between metric and density perturbations
is not so simple. Unfortunately, so far nobody has been able to address this problem in full generality and most attempts rely either
on approximations or on toy models. The latter are fully relativistic solutions which, however, impose symmetries which are not found
in the observed Universe.

In the past some of us have considered plane symmetric dust universes to study the effect of large overdensities on the
distance--redshift relation, $d_L(z)$, and on the Hubble parameter, $H(z)$~\cite{DDV} (see also \cite{Bull} for related work).
We found that even though the Hubble parameter
becomes strongly fluctuating in these solutions, in its integral, $d_L(z)$, the fluctuations average out and the deviations from
the Einstein-de~Sitter result are small. They can become somewhat larger for a line of sight parallel to the plane of symmetry along a sheet-like void, but they never exceed the luminosity distance for an empty (Milne) universe which up to about $z=1$ is still smaller than the observed distance $d_L(z)$, which is well fitted by the cosmological Lambda-cold-dark-matter standard model, $\La$CDM, dominated by a cosmological constant.

The problem of the exact plane symmetric dust solution is the fact that we have to choose relatively small initial density perturbations, otherwise we encounter a singularity before the present time. This singularity is a consequence of shell-crossing caustics which cannot be handled properly in a fluid approach and which become relevant as soon as density perturbations are large, which is exactly where we expect deviations of $d_L(z)$ from the FL relation to become relevant. This prompted us to study the problem with a method which can handle shell crossings but is still relativistic. This is what we attempt in this paper:

We re-examine the wall-universe scenario with a new approach, using the novel general relativistic $N$-body scheme
which was recently presented in \cite{ADDK}. 

In section~\ref{s:rel} we briefly summarize the key points of the two approaches, and in section~\ref{s:comp} we compare the relativistic $N$-body results with the exact solution
in the regime where both coexist. As soon as the evolution of dust leads to the formation of caustics, the exact solution
becomes singular due to the breakdown of the fluid description and of the synchronous gauge. Nothing serious happens though within the $N$-body scheme, which
employs longitudinal gauge.
It is therefore possible to study the solutions
in the highly non-linear regime beyond the formation of caustics, as is presented in section~\ref{s:sing}. We also comment on the comparison with traditional Newtonian $N$-body schemes in section~\ref{sec:results_longitudinal}.
We present our conclusions in section~\ref{s:con}.

\section{Relativistic and semi-relativistic wall universes}\label{s:rel}

\subsection{Description in synchronous comoving gauge}

We consider a model where the Universe contains only dust and is invariant under transformations (translations, rotations) of
the two-dimensional Euclidean group. In other words, two of the three space dimensions are homogeneous and isotropic. Perturbations
occur only in the form of plane-parallel sheets of over- and underdense regions. By construction, all perturbations (linear or
nonlinear) are confined to the scalar sector. In synchronous gauge, the metric takes the form
\be
\label{eq:metric-sync}
 ds^2 = -dt^2 + \al^2(t,\mathrm{x}) d\mathrm{x}^2 + \beta^2(t,\mathrm{x}) \left[d\mathrm{y}_1^2 + d\mathrm{y}_2^2\right]\, .
\ee

The Einstein equations for this geometry and for pure dust matter 
 yield~\cite{zakharov,szekeres,kras,DDV}
\bea
\dd_t\left(\frac{\beta'}{\al}\right) \equiv \dd_tE &=& 0 \label{e:0i}\, , \\
\left(\dd_t \beta\right)^2 - \left(\frac{\beta'}{\al}\right)^2 &=& 2   \frac{M(\mathrm{x})}{\beta} \, , \label{e:00}\\
M' = 4\pi G\rho \beta^2\beta' &=&  4\pi G\rho \beta^2\al E(\mathrm{x})  \label{e:M'}\,.
\eea
Here a prime denotes a derivative with respect to $\mathrm x$.
In Eq.~(\ref{e:0i}) we have introduced the time-independent function 
\be\label{e:E}
E(\mathrm{x}) = \beta'/\al
\ee
and Eq.~(\ref{e:00}) defines $M(\mathrm{x})$ which is also time-independent. 
Here we have assumed that matter is comoving. As long as the perfect fluid description is valid, this can always be achieved by a suitable choice of coordinates~\cite{kras}. We therefore call this the synchronous comoving gauge.
In Ref.~\cite{DDV} some of us have considered overdense ``walls'' separated by underdense regions of different sizes. There we also present the parametric solutions of Eqs.~(\ref{e:0i}) to (\ref{e:M'}) for a given initial density profile.
High overdensities turn the initial expansion of the fluid into contraction and rapidly lead to a shell-crossing singularity. In Ref.~\cite{DDV} we had to choose the initial overdensities such that no singularity was encountered up to the present time. In these Universes, the directions normal to the collapse, i.e.\ parallel to the wall, $(\mathrm{y}_1,\mathrm{y}_2)$ always expand.

\subsection{Description in longitudinal gauge}

The synchronous comoving gauge is useful as long as the dust can be described as a perfect fluid, in which case the evolution equations
reduce to a tractable set of differential equations as we have seen above. However, the fluid description usually breaks down during
nonlinear evolution due to the formation of caustics, i.e.\ the convergence of world lines of different fluid elements in the same
space-time point. This problem can be avoided by employing a particle description of the dust, which samples
the full phase space. In general we do not work at the level of the fundamental particles but instead use an $N$-body simulation
that samples the phase-space distribution by following the evolution of a relatively small set of ``representatives''.

In this approach, the equations become much more involved, and in general we are not able to find exact solutions but have to resort to numerical simulations.
$N$-body simulations traditionally employ Newton's laws of gravity rather than full General Relativity. There is, of course, at best an approximate
correspondence between Newtonian and relativistic cosmologies. On the formal level, this correspondence has been elaborated
in recent years \cite{GreenWald, ChisariZaldarriaga}, leading to a deeper understanding of the impressive success of Newtonian simulations within
the cosmological standard model. In general terms, this success rests upon the standard model assumptions that gravitational
fields are weak (on the relevant scales) and velocities are small. Furthermore, the effect of a cosmological constant can be taken
into account simply by adjusting the background.

Scenarios which potentially violate one of these assumptions can not be tested
reliably with Newtonian simulations. Examples include various models of dynamical DE or, to some extent, warm dark
matter. Some of us have therefore started to develop the numerical techniques for relativistic $N$-body simulations, which
incorporate a truly dynamical spacetime \cite{ADDK}. Since we want to remain within the cosmological context, we employ a weak-field
approximation which is described in detail in  \cite{ADDK} and which is closely related to the approach of \cite{GreenWald}. The equations are solved in longitudinal
gauge, in which metric perturbations indeed remain small on the relevant scales. For the plane-symmetric setup studied in this paper,
the metric in longitudinal gauge reads
\begin{multline}
\label{eq:metric-long}
 ds^2 = a^2(\tau) \left[-\left(1 + 2 \Psi(\tau, x^1)\right) d\tau^2\right. \\
 \left. + \left(1 - 2 \Phi(\tau, x^1)\right) \delta_{ij} dx^i dx^j\right] \, .
\end{multline}
In the scheme implemented numerically, we assume that the gravitational potentials $\Phi$ and $\Psi$ are small, of order $\ep$, but may
fluctuate on small spatial scales. This is taken into account by giving spatial derivatives a weight $\ep^{-1/2}$.
We then include all terms in Einstein's equations up to order $\ep$. In this scheme cold dark matter (CDM) velocities are of order $\ep^{1/2}$, while density fluctuations are large, of order $\ep^0$.
This scheme is fully relativistic up to the order described here. It just cannot handle 
large gravitational potentials, but these are not observed 
on cosmologically interesting scales $\la \gtrsim 0.1$ Mpc. Relativistic velocities $v \sim 1$ can be accommodated within the scheme, but since they do not arise, we simply truncate the stress-energy tensor and geodesic equation for CDM particles at order $v^2 \sim \ep$. More details are given in 
Ref.~\cite{ADDK}, where also the explicit equations can be found.

Here we identify the spacetime directions perpendicular to the plane of symmetry, $(t,\mathrm{x})$ in Eq.~(\ref{eq:metric-sync}), with the
$(\tau, x^1)$-plane. Note, however, that the \textit{coordinates} $\mathrm{x}$ and $x^1$ should not be identified. As long as the
perfect fluid description of dust is valid, the two metrics~(\ref{eq:metric-sync}) and (\ref{eq:metric-long}) are related
by a gauge transformation which eventually becomes nonlinear as matter perturbations grow, see section~\ref{s:trafo}. 

While the relativistic $N$-body approach does not have issues with shell crossing as it samples the full phase space, it relies
on the assumption that the metric perturbations remain small so that they can be treated perturbatively in a weak-field limit.
In section \ref{s:sing} we investigate whether this condition remains valid for the wall universes studied here.

\subsection{Transformations between the two gauges}\label{s:trafo}
At early times the Universe is close to Friedmann and we can write 
\be
\al(t,\mathrm{x}) = a(t)\left(1 + g(t,\mathrm{x}) \right)\, , ~
\beta(t,\mathrm{x}) = a(t)\left(1 + h(t,\mathrm{x}) \right)\,,
\ee
where $g$ and $h$ denote small perturbations. In conformal time $\tilde\tau$ defined by $d\tilde\tau = a^{-1}dt$ we obtain
\be
ds^2 = a^2(\tilde\tau)\left[-d\tilde\tau^2 + (1+2g)d\mathrm{x}^2 +  (1+2h)(d\mathrm{y}_1^2 +d\mathrm{y}_2^2)\right]
\ee
To determine the initial Bardeen potentials corresponding to these perturbations, we have to transform it to longitudinal gauge. We choose
$ \tilde\tau = \tau + T$ and $\mathrm{x} = x^1 +L$. In order for this to transform from synchronous to longitudinal gauge we must require 
$$g_{\rm long} = g_{\rm syn} +L_X\bar g\,, $$ where
$X=(T,L,0,0)$ is the vector field inducing the gauge transformation, $\bar g =a^2(-,+,+,+)$ is the background metric, $g_{\rm long}$ and $ g_{\rm syn}$ are the perturbed metric in longitudinal respectively synchronous gauge. $L_X$ denotes the Lie derivative in direction $X$. A brief calculation of the different terms yields 
\bea
\Psi +\HH T +\dot T &=0 \,,\quad\qquad \dot L - T' &= 0 \,,\\
-\Phi +\HH T + L' &=g \,,\qquad
-\Phi +\HH T &= h \,,\\
 \mbox { hence} && \nonumber\\
L' = g-h\,, & \quad T'' =\dot g - \dot h \,.
\eea
Here $\HH =(da/d\tau)a^{-1}$, and a dot denotes the derivative with respect to $\tau$. Hence $\Phi$ and $\Psi$ are the solutions of
\bea
\Phi'' = -h'' +\HH (\dot g - \dot h) \quad \mbox{and} \label{e:Phiin}\\
\Psi'' = \HH (\dot h - \dot g) +\ddot h - \ddot g \,. \label{e:Psiin}
\eea

For our simulations, we choose initial density profiles which deviate from a constant 
by the addition of a periodic plane-wave perturbation of small amplitude, see Ref.~\cite{DDV}.
We then determine the initial Bardeen potentials via Eqs.~(\ref{e:Phiin}), (\ref{e:Psiin}).
Note that the assumption of a flat FL background in Eq.~(\ref{eq:metric-long}) puts a constraint on the total matter density. This constraint finds its counterpart in Eqs.~(\ref{e:Phiin}), (\ref{e:Psiin}) by the requirement that they have appropriate periodic solutions. The perfect fluid solutions in comoving synchronous gauge introduced above are more general, allowing for arbitrary deviations from critical density. 
To study such general solutions within the $N$-body framework would require to extend the simulations to handle arbitrary background curvature.

At late times, the coordinate transformations are not simple linear gauge transformations. 
Finally, at shell-crossing, the synchronous, comoving gauge breaks down, while the longitudinal gauge is still well defined. In the fluid approximation, shell crossing corresponds to a real singularity in the density and therefore also in the curvature.
But since this singularity is sheet-like, the Christoffel symbols only have a jump and the metric components have a kink at the singularity. Such sheet-like singularities can be handled with the Israel junction method~\cite{Israel}. 
In the $N$-body approach such singular sheets do not lead to a divergent density on the grid; the finite lattice unit acts as a regulator. It should be noted that the singularity is introduced by the fluid approximation in the first place, and will be regulated in a similar fashion by fluid imperfection as soon as one considers a physical dark matter model.

\subsection{Observables}

Comparing quantities which have been calculated in different gauges can be quite subtle. The safest approach is to use observables,
as they are uniquely defined through a physical prescription, and therefore are gauge invariant by design. Two common
observables are redshifts and distances, which allow to construct the distance--redshift relation. It was this relation
that led to the discovery of the accelerated expansion of the Universe.

To determine the redshift and the distance to a source at some position 
$\bx_\ast$, we consider a photon emitted from the source at time $t_\ast$
arriving at the location of the observer, today, $(t_0, \bx_0)$. We denote the matter four-velocity
 field, hence the four-velocity of source and observer by $u(t, \bx)$ and the photon four-velocity 
 by $n$. The redshift of the source, $z$, is simply given by
 \be \label{e:redshift1}
 1+ z = \frac{g(n,u)|_\ast}{g(n,u)|_0} \,,
 \ee
where $g(n,u) \equiv g_{\mu\nu} n^\mu u^\nu$ denotes the usual scalar product induced by the metric. The evolution of the distance to the source is determined by the Sachs focussing equation~\cite{sachs,straumann},
\be\label{e:foc}
\frac{d^2d_A}{ds^2} = -\left(|\si|^2 + \RR\right)d_A \,.
\ee
Here $s$ is an affine parameter along the photon geodesic, and
\be
\RR = \frac{1}{2}R_{\mu\nu}n^\mu n^\nu = 4\pi GT_{\mu\nu}n^\mu n^\nu \,. 
\ee
The complex shear $\si$ of the light bundle is defined by
\be\label{e:shear}
\si = \frac{1}{2}g(\ep,\nabla_\ep n)\, , \mbox{ where }~ \ep \equiv e_1 + i e_2 \, .
\ee
The spatial orthonormal  vectors  $e_1$ and $e_2$ which are normal to both $u$ and $n$ 
at the observer are parallel transported along $n$, such that $\nabla_n e_j =0$.
They form a basis of the so called `screen'. For the explicit expressions see Appendix \ref{app:lumdist}.

The angular diameter distance $d_A$ to the source is defined as the solution of Eq.~(\ref{e:foc}) with final conditions
\be
\left. d_A\right|_0 = 0, \qquad \left. \frac{d \ d_A}{ds} \right|_0 = g(n,u)|_0.
\ee
The luminosity distance is related to the angular diameter distance via Etherington's reciprocity relation~\cite{Etherington}
$$ d_L(z) =(1+z)^2d_A(z) \,. $$

In the discussion of the results below we plot the distance modulus $\mu(z)$ (the $\log$ of the luminosity distance) and subtract the value one would obtain in the homogeneous model,
\be
\mu(z) - \mu_\mathrm{EdS}(z) = 5\log_{10}\left(\frac{d_L(z)}{d_L^{\rm EdS}(z)}\right)\,, 
\ee
 where $d_L^{\rm EdS}$ is the luminosity distance in an Einstein-de~Sitter Universe, i.e.\ in a matter dominated FL
 Universe with  vanishing curvature. We shall  compare our result to the distance in a Milne Universe, i.e.\ an empty FL Universe with negative curvature and to the standard $\La$CDM case. The expressions for the distances in these universes  are
\bean
 d_L^{\rm EdS}(z) &=&\frac{2}{H_0}\left(1+z -\sqrt{1+z}\right)\,, \\
 d_L^{\rm Milne}(z) &=&\frac{1}{H_0}\left(z+\frac{z^2}{2}\right)\, , \\
 d_L^{\La\rm{CDM}}(z) &=& \frac{1+z}{H_0}\int_0^z \frac{dz'}{\sqrt{\Om_m(1+z')^3 +\Om_\La}} \,.
\eean

A second observable is given by the image distortion which is induced by the shear alone. This distortion can be measured with the help of weak lensing observations. It can be characterized by a complex quantity $e$ whose absolute value measures the ellipticity acquired by an infinitesimal light bundle with circular cross-section at the observer when traced back along the photon path \cite{Perlick}. The phase angle of $e$, on the other hand, encodes the orientation of the principal axis of the elliptical cross section with respect to the screen vectors. This complex quantity is obtained by integrating the shear according to
\be \label{eq:ellipticity}
\frac{d e}{d s} = 2 \sigma \sqrt{|e|^2 + 4} \, ,
\ee
with final condition $e|_0 = 0$.

We will plot the absolute value of $e$ (in cases where it is not zero by symmetry) as a function of observed redshift. The direct interpretation of this quantity is the ellipticity of the observed image of a source with intrinsically circular shape, located at the given redshift. If the ratio of the principle axes of the observed elliptical image is $r$ then $|e|=|r-(1/r)|$, so that $r=2$ corresponds to $|e|=3/2$.
We will not plot the phase angle of $e$, because it is already fully determined by the symmetry of our setup.
 
\section{Comparison and interpretation of the results}

\subsection{Exact relativistic and $N$-body solutions}\label{s:comp}

\begin{figure}[t]
 \includegraphics[width=\columnwidth]{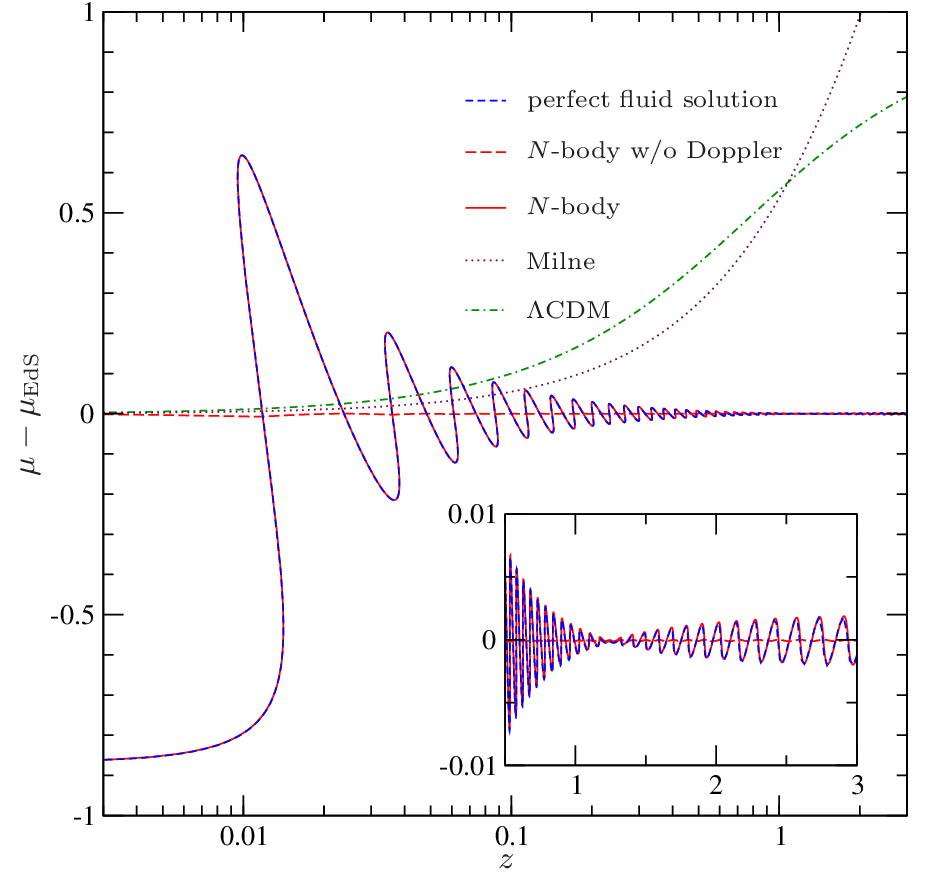}
 \caption{\label{fig:dmu} 
 (Color online) We plot the observed distance--redshift relation for a plane symmetric setup with an initial plane-wave perturbation of comoving wavelength $\lambda = 70$ Mpc$/h$. More precisely, we
plot the distance modulus minus the one of an Einstein-de~Sitter background. We compare the perfect fluid description (blue, dashed) with a relativistic $N$-body simulation with the same initial conditions (red, solid). The observer sits at the center of an underdense region and the line of sight is chosen perpendicular to the plane of symmetry. In this case, within the longitudinal gauge employed in the $N$-body framework, the fluctuations of the distance are dominated by the peculiar motion (Doppler) of the sources, which are assumed to follow the bulk flow of CDM. To illustrate this, we also plot the luminosity distance with the Doppler term subtracted (red, long-dashed). For a better assessment of the size of the effects, we also indicate  the distance--redshift relation for two well-known FL models: the Milne universe (purple, dotted) and a $\Lambda$CDM model with $\Omega_\Lambda = 2/3$ (green, dot-dashed).
The inset shows a zoom into the region $z > 0.5$ and has a linear $z$-axis}
\end{figure}

In Fig.\ \ref{fig:dmu} we show the distance--redshift relation for a plane wave perturbation initially described by a cosine-function with a comoving wavelength of 70 Mpc/$h$, for an observer located at the center of the underdense region and a photon coming in perpendicular to the plane of symmetry. We show both the exact
solution for a perfect dust fluid (blue, dashed) and the relativistic $N$-body simulation (red, solid). The two distances agree
extremely well over the entire redshift range
which provides an important check of the accuracy of the relativistic $N$-body approach.

\begin{figure}[t]
\includegraphics[width=\columnwidth]{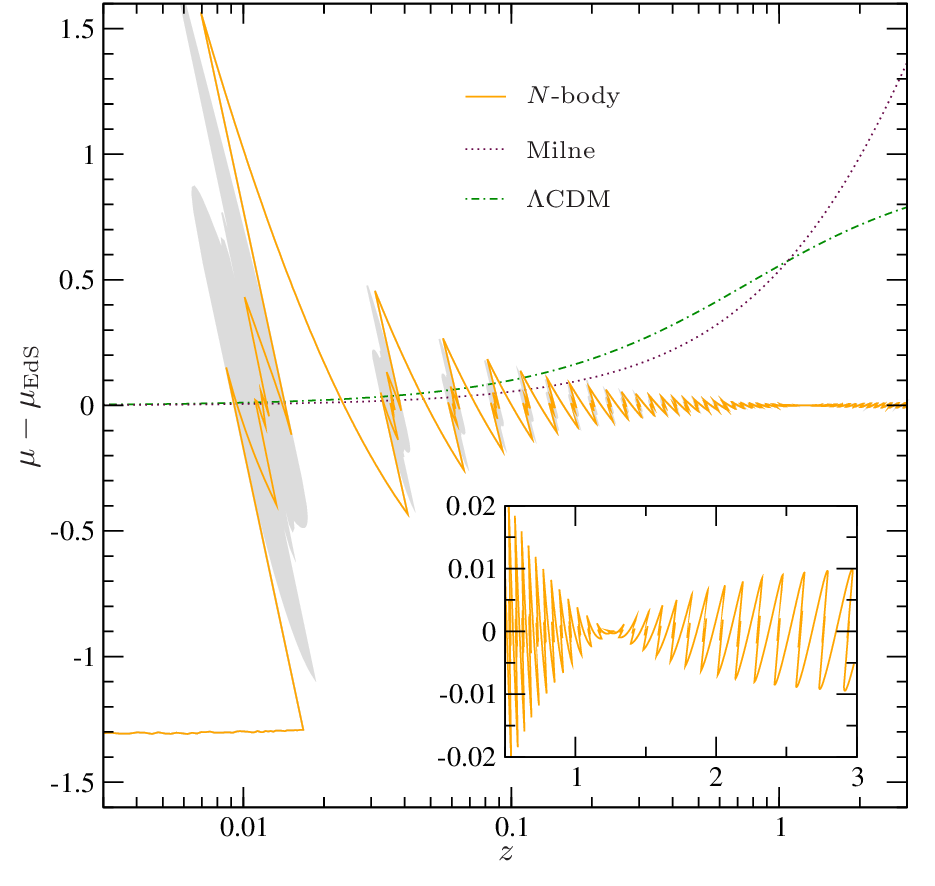}
\caption{\label{fig:dmu2}
(Color online) Same as Fig.~\ref{fig:dmu}, but with an initial perturbation amplitude which is five times larger. In this case the
  perfect fluid description breaks down due to the formation of caustics before the end of the simulation. We therefore only show the $N$-body result (orange, solid). In the regions where shell-crossing has occurred, the phase space distribution of particles shows a considerable velocity dispersion (see Fig.~\ref{fig:phasespace}). One may expect a similar dispersion for the sources: the gray areas correspond to the possible observed scatter of the distance modulus induced by the typical standard deviation of source velocities with respect to the bulk flow (at $1 \sigma$).}
\end{figure}
\begin{figure}[t]
\includegraphics[width=\columnwidth]{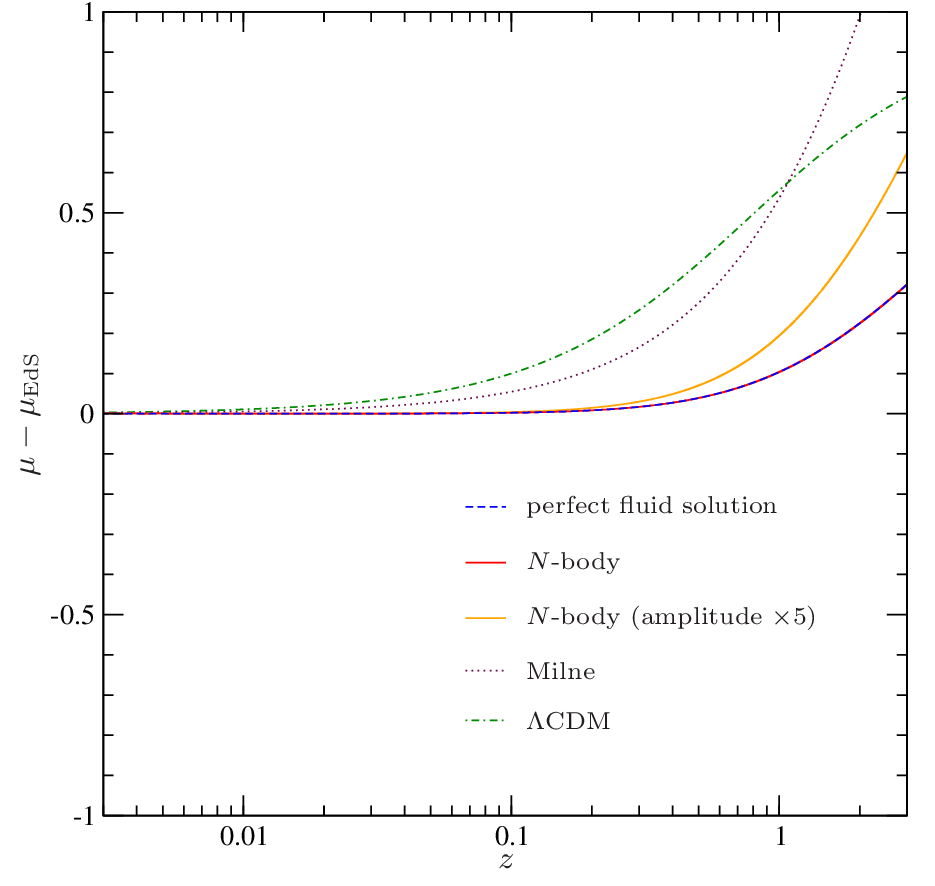}
\caption{\label{fig:dmu3}
(Color online) Same as Figs.~\ref{fig:dmu} and \ref{fig:dmu2}, but with the line of sight parallel to the plane of symmetry. In this case, the photon trajectory runs within the center of an underdense sheet, and the peculiar motion of sources is zero by construction also in  longitudinal gauge. However, since the beam is continuously defocussed due to the underdensity, a considerable deviation from the unperturbed distance--redshift relation is accumulated. The deviation remains below the one for the Milne universe and would be too small to account for DE.}
\end{figure}

\begin{figure}[t]
\includegraphics[width=\columnwidth]{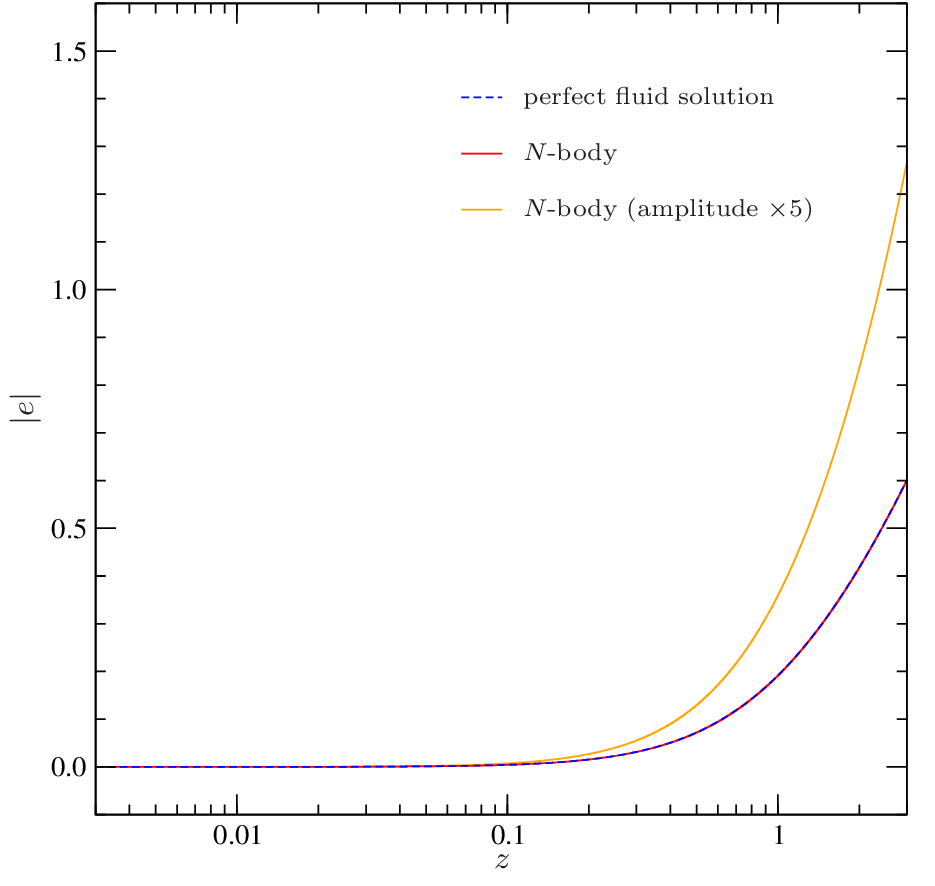}
\caption{\label{fig:e}
(Color online) Image distortion, measured by the ellipticity $|e|$ observed for small circular sources with redshift $z$. We consider the same scenarios as in Fig.~\ref{fig:dmu3}. An ellipticity of $|e|=3/2$ implies that one principle axis is twice as long as the other.}
\end{figure}

Comparing the result to Milne (purple, dotted)  or $\La$CDM (green, dot-dashed), we find that for $z \gtrsim 0.2$ the deviations from EdS are much too small to mimic observations which are in good agreement with $\La$CDM. They also have the wrong shape. It is however possible that this changes when we allow for larger initial perturbations that become non-linear and undergo shell-crossing before the present time. This situation can only be modeled in the $N$-body approach, and we show the result in Fig.\ \ref{fig:dmu2}, where we used an initial perturbation that is five times larger than the one shown in Fig.\ \ref{fig:dmu}. In this case the fluctuations are indeed larger by a factor of about 2, but at high redshift ($z\gtrsim 0.3$) they are still too small, and again they have the wrong shape.

As discussed in more detail below, the most important contribution to the fluctuations in the luminosity distance comes from the Doppler effect, i.e.\ the peculiar velocity of the object emitting the light. This is visible in Fig.\ \ref{fig:dmu}, where we also plot the luminosity distance without this term (red dashed curve). Once shell-crossing has occurred, there is no longer a single, well-defined velocity field at each point in space. Instead we now have a velocity dispersion. As the Doppler effect is so important, one might hope that taking into account velocity dispersion could help yield a better fit to the $\Lambda$CDM curve. 
But as shown by the gray area in Fig.\ \ref{fig:dmu2}, which represents the velocity dispersion, this is not the case. Velocity dispersion is mainly relevant at low redshift, $z<0.1$, while the deviation from $\Lambda$CDM is most significant at larger redshift.

We also see that the luminosity distance is in general not single valued (as pointed out previously e.g.\ in \cite{Mustapha:1997xb}). It seems intriguing that
the impact of the fluctuations on the distance is minimal around $z = 1.25$, cf.\ the inset of Figs.\ \ref{fig:dmu} and \ref{fig:dmu2}, which happens to be also the redshift at which the angular diameter distance in the Einstein-de~Sitter model is maximal. However, in the longitudinal gauge, this can be easily understood from the fact that the main effect (in these particular cases) is caused by peculiar motions: a perturbation of the redshift (as caused by the Doppler effect) changes our observable $\mu(z) - \mu_\mathrm{EdS}(z)$ at first order by $\delta z \times d \ln d_A^\mathrm{EdS} / dz$. This expression becomes zero at the maximum of $d_A^\mathrm{EdS}$.

Instead of looking at photons that propagate in the direction perpendicular to the plane of symmetry (i.e.\ `across' the perturbations), we can also consider photons that move along the symmetry directions, i.e.\ that follow a trough of the density along the $\mathrm{y}$-direction. In this case the lack of matter along the photon path leads to a continuous defocussing of the light beam which is only slightly counteracted by the presence of a non-zero shear (generated through the Weyl tensor). However, as shown in Fig.\ \ref{fig:dmu3}, this increase of the luminosity distance is still not sufficient to mimic $\Lambda$CDM. In fact, the luminosity distance remains strictly smaller than the one of the Milne Universe.

The non-zero complex shear generates an ellipticity for light bundles propagating along the symmetry directions, and we plot the absolute value of the ellipticity as a function of observed redshift in Fig.\ \ref{fig:e}. As this is a cumulative effect, the ellipticity can become very
large at redshifts of order unity.

\begin{figure}
\includegraphics[width=\columnwidth]{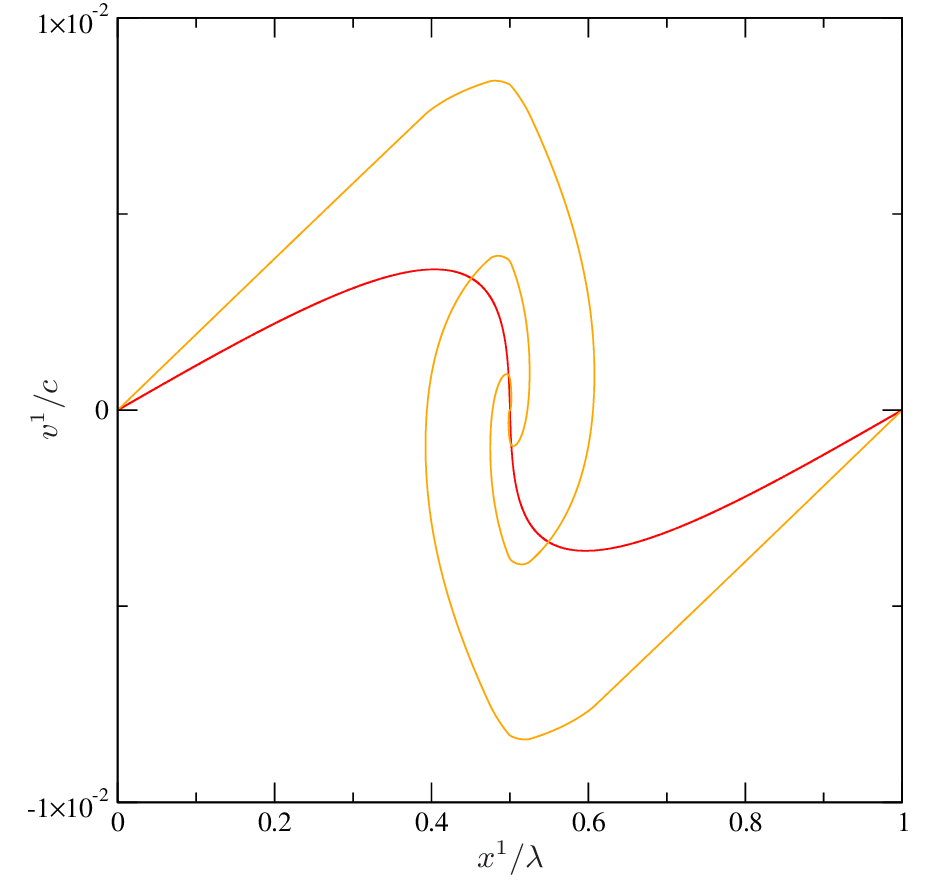}
\caption{\label{fig:phasespace}
(Color online) Phase space diagram at the end of the simulation for the setup used for Fig.\ \ref{fig:dmu} (red, inner curve) and
Fig.\ \ref{fig:dmu2} (orange, outer curve). In the first case shell-crossing does not occur and the velocity is single-valued everywhere. In the second case on the other hand shell-crossing has occurred, and near the central overdensity the velocity field is multi-valued. In this case there is a velocity dispersion that leads to a dispersion also in the luminosity distance, see Fig.\ \ref{fig:dmu2}.}
\end{figure}

\subsection{Shell-crossing and singularities}\label{s:sing}

The orange solid curve in Figs.\ \ref{fig:dmu2} and \ref{fig:dmu3} shows the distance-redshift relation in the  more extreme setup where shell-crossing occurs before today.
This section investigates in more detail what happens in this situation.

A caustic formally leads to a divergent stress-energy tensor. The divergence, however, occurs
in the form of a delta-function on a $(2+1)$-dimensional (timelike) worldsheet and can in principle be handled with the Israel
junction method~\cite{Israel}. The particle acceleration remains small everywhere but is discontinuous on the caustic. In the $N$-body treatment, the
discontinuity is smeared out by the finite spatial dispersion of the $N$-body particles. This is similar to physical reality,
but on the scale of the  $N$-body particles which are much larger than the true microscopic CDM particles.

\begin{figure}
\includegraphics[width=\columnwidth]{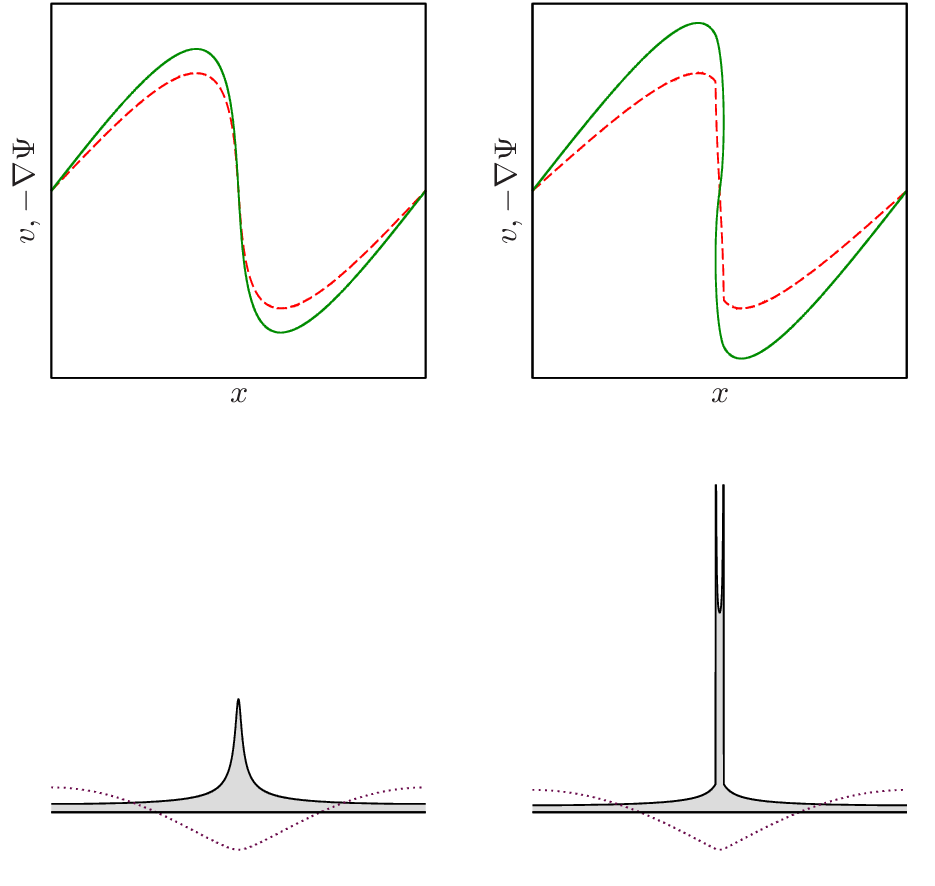}
\caption{\label{fig:shellcrossing}
(Color online) Sketch of the situation shortly before and after shell crossing. The upper panels show the phase space distribution (green, solid) and the local acceleration, given by the gradient of $\Psi$ (red, dashed). The lower panels show the density (black, solid) and the potential $\Psi$ itself (purple, dotted) -- the potential $\Phi$ is for all practical purposes indistinguishable from $\Psi$ for the cases studied here. On the left side, shell crossing has not yet occurred: the perfect fluid description (zero velocity dispersion) is still valid and all quantities are regular and smooth. On the right side, two caustics have formed as a result of shell crossing. The density on the caustics diverges like a delta-function on a sheet. However, the potential itself remains small everywhere. Its gradient, which corresponds to the acceleration, also remains small, but it is discontinuous on the caustics. Only the second derivative of $\Psi$ becomes large.}
\end{figure}

In the phase space representation at the present time given in  Fig.~\ref{fig:phasespace} we see that shell-crossing has happened by today in this setup (orange, outer line) while it is just about to happen in the setup with 5 times smaller initial density contrast (red, inner line).
In Fig.~\ref{fig:shellcrossing} we show, in addition to the phase space, also the acceleration, the matter density and the gravitational potential just before and just after shell crossing happens. We can observe the jump in the acceleration (red-dashed)
after shell crossing occurs (right panel) which is induced by a kink in the gravitational potential $\Psi$ (purple dotted in the lower panel). The only quantity that becomes large at shell crossing is the density.

The comoving synchronous gauge becomes singular at shell crossing, since the fluid rest frame is no longer well-defined. It is remarkable that the longitudinal gauge does not
only remain finite, but the metric potentials stay small and safely in the perturbative regime. Therefore, our relativistic $N$-body simulation in longitudinal gauge is well adapted to describe non-linear structure formation.

\subsection{Interpretation in longitudinal gauge \label{sec:results_longitudinal}}

When we compare the result of the relativistic $N$-body simulation with a Newtonian $N$-body simulation
for the scenario shown in Fig.\ \ref{fig:dmu}, we find that the two agree extremely well if we use the relativistic
distance formula including the Doppler term for both cases. The reason is that in  longitudinal
gauge the result is entirely dominated by the contributions from peculiar velocities, see Fig.\ \ref{fig:dmu}. The contributions from the metric potentials are sub-dominant, 
and  vector and tensor perturbations are absent in an exactly plane-symmetric universe. We do expect velocities which are of order $\ep^{1/2}$ to dominate over the gravitational potentials which are of order $\ep$ in our counting scheme.

This dominance of the Doppler contribution is however gauge- and situation-dependent. In the synchronous co-moving gauge the particles are by
definition at rest with respect to the coordinate system. If we look along a symmetric direction $\mathrm{y}$ rather than in the transverse direction $\mathrm{x}$, the velocities
are zero also in longitudinal gauge. In this case, perturbations of the distance modulus are entirely governed by the gravitational potentials.

\section{Conclusions}\label{s:con}

In this paper we have analyzed the distance--redshift relation for plane symmetric universes.
We have used general relativistic fluid solutions as well as relativistic $N$-body simulations.  
The fluid approach suffers from singularities due to the formation of caustics and cannot be used for high overdensities: when particle trajectories cross, the comoving synchronous coordinate system used for these exact solutions breaks down. The $N$-body approach, which is a phase-space method, remains regular at all times. 
Furthermore, in this approach the gravitational potentials remain small so that our approximation is consistent.  

Both approaches give consistent results where they are both regular: wall inhomogeneities, even though they do modify the distance--redshift relation, cannot mimic Dark Energy. This is also true for high density fluctuations where the fluid approach breaks down and which is treated with our relativistic $N$-body code developed in Ref~\cite{ADDK}. The excellent agreement with the exact relativistic solution in the scenarios without shell crossing provides an important test for the accuracy of our code.

It is well known that inhomogeneous models can in principle reproduce any given distance--redshift relation by carefully adjusting the matter distribution. Such fine-tuned models typically violate the cosmological principle. In a model where the Universe has a homogeneity scale well within the observed patch (which is the only case where one can talk about the cosmological principle being respected), the distance--redshift relation can still be affected by the lumpiness of matter. However, fluctuations induced by peculiar motion are expected to average out when considering a large enough sample of observations. Defocussing of light beams, which for realistic observations are biased towards travelling mostly through underdense regions, does not seem to give a strong enough effect to be mistaken for Dark Energy. In fact, we were not even able to reach the level of defocussing found in the Milne model, even when considering light beams which travelled through essentially depleted regions only.

If the results from this toy model can be generalized, we have to conclude that structures in the Universe -- over- and underdensities -- cannot be responsible for the observed acceleration and Dark Energy or Dark Gravity is needed to explain it. Nevertheless, they certainly do affect the distance--redshift relation and therefore they have to be taken into account when interpreting measurements precisely.

\begin{acknowledgments}
It is a pleasure to thank Philip Bull  for interesting discussions.
J.A.\ acknowledges funding from the German Research Foundation (DFG) through the research fellowship \mbox{AD~439/1-1}. This
work is supported  by the Swiss National Science Foundation. Some of the simulations for this paper used the Andromeda cluster of the University of Geneva.
\end{acknowledgments}

\appendix
\section{Computing the distance redshift relation\label{app:lumdist}}
In this appendix we explain in more detail how we calculate the luminosity distance in the $N$-body code. Details for the exact relativistic fluid solution can be found in Ref.~\cite{DDV}.

The relativistic $N$-body scheme we use has the advantage that the metric is explicitly computed, and one can therefore directly integrate the null geodesic equation numerically to obtain the path of a photon in the perturbed geometry. The difficulty, when constructing observables like the distance--redshift relation,  is that they are defined on the past light cone of an observer, whereas the simulation, on the other hand, evolves forward in time. In the perturbed Universe, one does not know exactly whether a point lies on the past light cone of an event until one has actually found a photon path which connects the two. Naturally, one would do this construction backwards in time, starting form the observer, but for this one needs to save a part of the four-dimensional geometry with high resolution in space and time. Although this is a possible way to go, we chose a different approach which one may call a ``shooting method''.

Somewhere close to the highest redshift which we want to plot in the distance--redshift diagram, we choose an initial point located at spacetime coordinates which would be connected to the observer event by a null ray in the \textit{unperturbed} geometry. This location can simply be read off from the distance--redshift relation obtained in an exact FL universe. We then shoot a light ray directed at the observer by integrating the null geodesic equation in the perturbed geometry, along with the $N$-body simulation. When the simulation reaches the observer event, we usually find that we have missed the observer by some small spacelike distance due to the perturbations of the photon path. We then restart the simulation, correcting the coordinates of the initial point by the amount by which we have missed the observer. Rerunning the simulation will now bring the perturbed light ray almost to the observer event, up to second-order perturbations. This procedure can be iterated to close in on the observer event 
to arbitrary precision. For our purposes, a single iteration was enough.

At the last iteration, we also save the change of the photon energy and some information about the geometry, in particular the terms which enter the Sachs equation, along the light ray with high resolution. Additionally, we also save the peculiar velocities of sources which lie on the path. Since baryonic physics is completely neglected in our simple setup, we assume that observable sources have the same distribution as CDM particles. With this set of data, once we have reached the end of the simulation, we can integrate the Sachs equation backwards in time along the line of sight given by the light ray. Eliminating the affine parameter $s$ in the equations in favor of the coordinate time $\tau$, the angular distance evolves according to

\begin{equation}
\label{eq:dA}
 \ddot{d_A} + \frac{\dot{n^0}}{n^0} \dot{d_A} + \left(|\sigma|^2 + \mathcal{R}\right)\frac{d_A}{\left(n^0\right)^2} = 0 \, ,
\end{equation}
where $n^0 = d\tau/ds$ is the $\tau$-component of the photon null vector which has to be determined from the null geodesic equation,
\begin{equation}
\label{eq:geodesic}
 \dot{n^0} + \left(\dot{\Phi} - \dot{\Psi} + 2 \nabla_\mathbf{n} \Psi + 2 \mathcal{H}\right) n^0 = 0 \, ,
\end{equation}
where $\mathbf{n}$ denotes the spatial direction of the photon vector. This equation takes care of the path-dependent contribution to the redshift. The total redshift (the ratio between photon energies measured in the rest frames of source and observer) is given by Eq.~(\ref{e:redshift1}), which yields

\begin{equation}
 1 + z = \frac{\left. \left[ n^0 a \left(1 + \Psi - \mathbf{n} \cdot \mathbf{v} + \frac{\mathbf{v}^2}{2} \right) \right] \right|_\ast}{\left. \left[ n^0 a \left(1 + \Psi - \mathbf{n} \cdot \mathbf{v} + \frac{\mathbf{v}^2}{2}\right) \right] \right|_0} \, .
\end{equation}
Here, $\mathbf{v}$ is the peculiar velocity vector in the longitudinal gauge, $\mathbf{n}\cdot\mathbf{v}$ denotes its projection on the photon direction, and the subscripts $\ast$ and $0$ indicate that the entire expression has to be evaluated at the source and observer event, respectively. The last two equations have been truncated at our approximation order.

We finally need an evolution equation for the complex shear, see~\cite{sachs,straumann}. For the purpose of solving Eq.~(\ref{eq:dA}), it is useful to write it as
\begin{equation}
\label{eq:shear}
 \frac{d}{d\tau}\left(\frac{\sigma}{n^0}\right) + \left(\frac{\dot{n^0}}{n^0} + 2 \frac{\dot{d_A}}{d_A}\right) \frac{\sigma}{n^0} + \frac{\FF}{\left(n^0\right)^2} = 0 \, ,
\end{equation}
where $\FF = \frac{1}{2} C_{\kappa\lambda\mu\nu} \epsilon^\kappa n^\lambda \epsilon^\mu n^\nu$ is a contraction of the Weyl tensor with the complex screen vector $\epsilon$ and the photon four velocity.

In order to solve this coupled system of differential equations, it is sufficient to know four real-valued quantities along the line of sight (as a function of $\tau$) which can all be obtained from the knowledge of the metric and the photon direction: $\dot{n^0} / n^0$, $\mathcal{R} / (n^0)^2$, and $\FF / (n^0)^2$. The last quantity is complex in general and therefore corresponds to two real-valued quantities. However, for the particular lines of sight we chose to study in this work, owing to the symmetry of our setup, we can choose the screen vectors such that $\FF / (n^0)^2$ remains real-valued. In particular, for the light ray perpendicular to the plane of symmetry, we find
\begin{eqnarray}
 \frac{\mathcal{R}}{\left(n^0\right)^2} &=& 4 \pi G a^2 \left[-\left(1 + 2 \Psi\right) T^0_0 + 2 T^1_0 + T^1_1\right] \, ,\nonumber\\
 \frac{\FF}{\left(n^0\right)^2} &=& 0 \, ,
\end{eqnarray}
up to terms which are neglected in our approximation scheme. In this case, the shear remains zero and the angular distance is entirely governed by convergence. For the light ray parallel to the plane of symmetry, we can choose an orthogonal basis of the screen where one basis vector remains orthogonal to the plane of symmetry, while the other basis vector remains parallel to it. This is possible because the center of the underdense region has an additional $\mathbb{Z}_2$ symmetry which guarantees that the ray remains parallel to the plane of symmetry (even though this path is unstable under small perturbations). Using such a basis, we find
\begin{eqnarray}
 \frac{\mathcal{R}}{\left(n^0\right)^2} &=& -4 \pi G a^2 \left(1 + 2 \Psi\right) T^0_0 \, ,\nonumber\\
 \frac{\FF}{\left(n^0\right)^2} &=& \left(\frac{1}{2} + 2 \Phi + \Psi\right) \Delta \Phi + \left(\frac{1}{2} + \Phi\right) \Delta\Psi \, .\qquad
\end{eqnarray}
In all these explicit expressions we have used the symmetries of our setup to simplify them. Note also that we assume $T^1_0$ is of order $\epsilon^{1/2}$ and $T^1_1$ is of order $\epsilon$; components with spatial indices $2$, $3$ vanish by symmetry.
Using this information, the distance--redshift relation is constructed by integrating Eqs.~(\ref{eq:dA}), (\ref{eq:geodesic}), (\ref{eq:shear}) backwards in time. To this end, the final conditions are fixed at the observer as $d_A(\tau_0) = 0$, $\dot{d_A}(\tau_0) = -a (1 + \Psi - \mathbf{n} \cdot \mathbf{v} + \mathbf{v}^2/2) |_0$, $\sigma(\tau_0) = 0$, and $n^0(\tau_0) > 0$ (arbitrary). The solution for the ellipticity $e$ in terms of the real-valued shear follows from Eq.~(\ref{eq:ellipticity}):
\begin{equation}
 e = 2 \sinh \left(2 \int_{\tau_0}^{\tau_\ast} \! \frac{\sigma}{n^0} d\tau\right) \, .
\end{equation}

\end{document}